\def\({ \left( }
\def\){ \right) }
\def\b{\begin{equation}}
\def\e{\end{equation}}
\def\={\ =\ }

\def\+{\ +\ }
\def\-{\ -\ }

\def\Ls{\cal L \rm}

\def\mumu{$\mu^+\mu^-$}
\def\ee{$e^+e^-$}

\documentstyle[linac96,times,epsfig]{article} 

\begin{document}
\title{HIGH LUMINOSITY MUON COLLIDER DESIGN}
\author{Robert Palmer, Juan Gallardo,\\
Brookhaven National Laboratory\\
  Upton, NY 11973-5000, USA}

\maketitle

%
\begin{abstract}
Muon Colliders have unique technical and physics advantages 
and disadvantages when compared with both hadron and 
electron machines. They should be regarded as 
complementary. Parameters are given of 4 TeV high 
luminosity \mumu collider, and of a 0.5 TeV lower 
luminosity demonstration machine. We discuss the various 
systems in such muon colliders. 
 \end{abstract}
 
\section{Introduction}

The possibility of muon colliders was introduced  by 
Skrinsky et
al.\cite{ref2}, Neuffer\cite{ref3},  and others. More 
recently, several
workshops and collaboration meetings have greatly 
increased the level of
discussion\cite{ref4},\cite{ref5}. A detailed 
Feasibility Study\cite{snow} was 
presented at Snowmass 96.

\subsection*{Technical Questions}

Hadron collider energies are limited by their size, and 
technical constraints on bending magnetic fields. Lepton 
(\ee or \mumu) colliders, because they undergo simple, 
single-particle interactions, can reach higher energy final 
states than an equivalent hadron machine. However, 
extension of $e^+e^-$  colliders to multi-TeV energies is 
severely performance-constrained by beamstrahlung, The 
luminosity $\Ls$ of a lepton collider can be written: 
 \b
   \label{lumeq}
\Ls \= {1 \over 4\pi E}\ \ {n_\gamma \over 2 r_o \alpha}
  \ \ {P_{beam} \over \sigma_y} \ \ n_{collisions}
\e
where $\sigma_y$ is the average vertical (assumed smaller) beam 
spot size, $E$ is the beam energy, $P_{beam}$ is the total 
beam power, $\alpha$ is the electromagnetic constant, $r_o$ 
is the classical radius, 
and $n_{\gamma}$ is the number of photons 
emitted by one bunch as it passes through the opposite one. 
If this number is too large then the beamstrahlung 
background of electron pairs and other products becomes 
unacceptable. 

    As the energy rises, the luminosity, for the same event 
rate, must rise as the square of the energy. For an electron 
collider, $n_{collisions}=1$, and, for a fixed background, 
we have the severe requirement: 
 \b
{P_{beam} \over \sigma_y}\ \propto \ E^3
 \e

   In a muon collider there are two significant changes: 1) 
The classical radius $r_o$ is now that for the muon and is 
200 times smaller; and 2) the number of collisions a bunch 
can make $n_{collisions}$ is no longer 1, but is now 
related to the average bending field in the muon collider 
ring, For 6~T, it is 900.

   In addition, with muons, synchrotron 
radiation is negligible, and the collider is circular. 
In practice this means that it can be much smaller than a 
linear electron machine. The linacs for the 0.5 TeV NLC 
will be 20 km long. The ring for a muon collider of the 
same energy would be only about 1.2 km circumference. 
            
  There are, of course, technical difficulties in making 
sufficient muons, cooling and accelerating them before they 
decay and dealing with the decay products in the collider 
ring. Despite these difficulties, it appears possible that 
high energy muon colliders might have luminosities 
comparable to or, at energies of several TeV, even higher 
than those in \ee colliders. 

\subsection{Parameters}

\begin{figure}[t!] 
\centerline{\epsfig{file=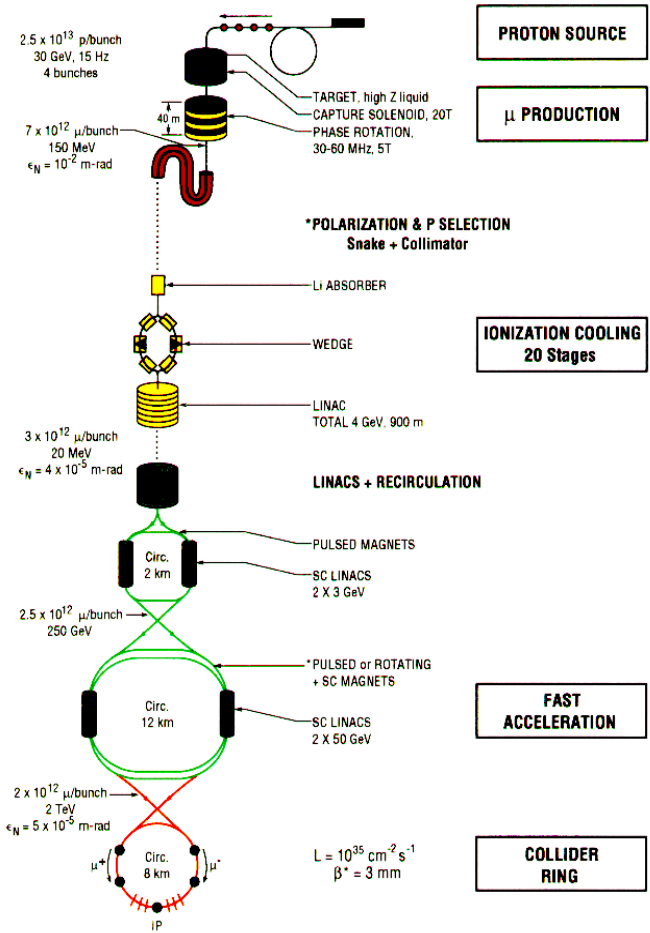,height=5.0in,width=3.5in}}
\caption{Schematic of a Muon Collider.}
\label{schematic}
\end{figure}

The basic parameters of a 4 TeV \mumu collider are shown 
schematically in Fig.\ref{schematic} and given in 
Tb.\ref{sum} together with those for a 0.5 TeV 
demonstration machine based on the AGS as an injector. It 
is assumed that a demonstration version based on upgrades 
of the FERMILAB machines would also be possible. 

\begin{table}[thb]  
\begin{tabular}{llcc}
C of m Energy              & TeV      & 4     & .5       
\\
Beam energy                & TeV      &     2    &   .25 
   \\
Beam $\gamma$              &          &   19,000 &  
2,400   \\
Repetition rate            & Hz       &    15    &    
2.5    \\
Muons per bunch            & $10^{12}$  &   2    &    4  
    \\
Bunches of each sign       &          &   2      &    1  
    \\
Norm. {\it rms} emit.   $\epsilon_N$   &$\pi$ mm mrad  & 
 50  &  90    \\
Bending Field              &  T   &    9    &    9     
\\
Circumference              &  Km      &    7    &    1.2 
   \\
Ave. ring  field $B$    & T   & 6    &   5       \\
Effective turns          &       & 900    &   800   \\
$\beta^*$ at intersection   & mm     &   3   &   8     
\\
{\it rms} I.P. beam size        & $\mu m$&   2.8 &  17   
 \\
Luminosity &${\rm cm}^{-2}{\rm s}^{-1}$& 
$10^{35}$&$10^{33}$\\
\end{tabular}
\caption{Parameters of Collider Rings}
\label{sum}
\end{table}

\section{Components}
\subsection{Proton Driver}

The proton driver is a high-intensity (four bunches of 
$2.5\times  10^{13}$ protons per pulse) 30 GeV proton 
synchrotron, operating at a repetition rate of 15 Hz. Two 
of the bunches are used to make $\mu^+$'s and two to make 
$\mu^-$'s. Prior to targeting the bunches are compressed to 
an rms length of 1 ns. 

For a demonstration machine using the AGS\cite{roser}, two 
bunches of $5\times  10^{13}$ at a repetition rate of 2.5 
Hz at $24\,$GeV could be used. 

\subsection{Target} 
Predictions of nuclear Monte-Carlo 
programs\cite{arc}\cite{MARS}\cite{other} suggest that $\pi$ 
production is maximized by the use of heavy target 
materials, and that the production is peaked at a 
relatively low pion energy ($\approx 100\,$MeV), 
substantially independent of the initial proton energy. 

Cooling requirements dictate that the target be liquid: 
liquid lead and 
gallium are under consideration. In order to avoid shock 
damage to a 
container, the liquid could be in the form of a jet.

\subsection*{Pion Capture}
 
Pions are captured from the target by a high-field 
($20\,$T, 15 cm aperture) hybrid magnet: 
superconducting on the outside, and a water cooled Bitter 
solenoid on the inside.   A preliminary design\cite{weggel} 
has a Bitter magnet with an inside coil diameter of 24 cm (space 
is allowed for a 4 cm heavy metal shield inside the coil) 
and an outside diameter of 60 cm; it provides half (10T) of 
the total field, and would consume approximately 8 MW. The 
superconducting magnet has a set of three coils, all with 
inside diameters of 70 cm and is designed to give 10 T at 
the target and provide the required tapered field to match 
into the decay channel. 

\subsection{Decay Channel and Phase Rotation Linac}

The decay channel consists of a periodic superconducting 
solenoidal ($5\,$T and radius $=15\,$cm).  A linac is 
introduced along the decay channel, with frequencies and 
phases chosen to deaccelerate the fast particles and 
accelerate the slow ones; i.e. to phase rotate the muon 
bunch.

\begin{figure}[hbt] 
\centerline{\epsfig{file=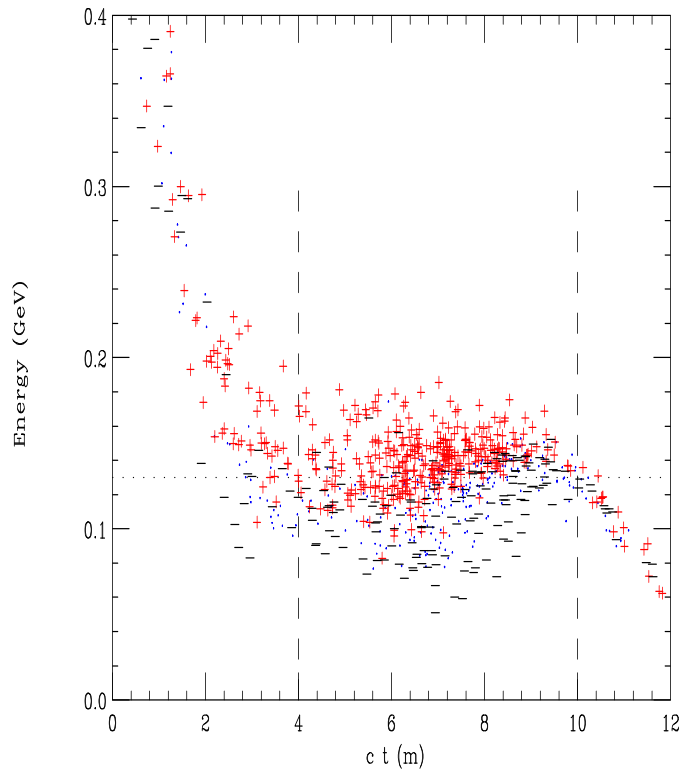,height=4.0in,width=3.5in}} 
\caption[Energy vs ct of muons at end of decay 
channel with phase rotation] {Energy vs ct of muons at end 
of decay channel with phase rotation; muons with 
polarization P$>{1\over 3}$, $-{1\over 3}< P<{1\over 3}$, 
and P$<-{ 1\over 3}$ are marked by the symbols `+', `.' and 
`-' respectively.   } 
 \label{Evsctpol2}
 \end{figure}

  Fig. \ref{Evsctpol2}  shows the energy vs ct at the end 
of the decay channel. 

The selected muons have a mean energy 150 MeV,
rms bunch length $1.7\,$m, and rms momentum  spread  
$20\,$\% ($95\,$\%,
$\epsilon_{\rm L}= 3.2\,{\rm eV s}$). The number of
muons per initial proton in this selected  bunch is 
$\approx$ 0.3.

\subsection*{Polarization Selection}

\begin{figure}[bht!] 
\centerline{\epsfig{file=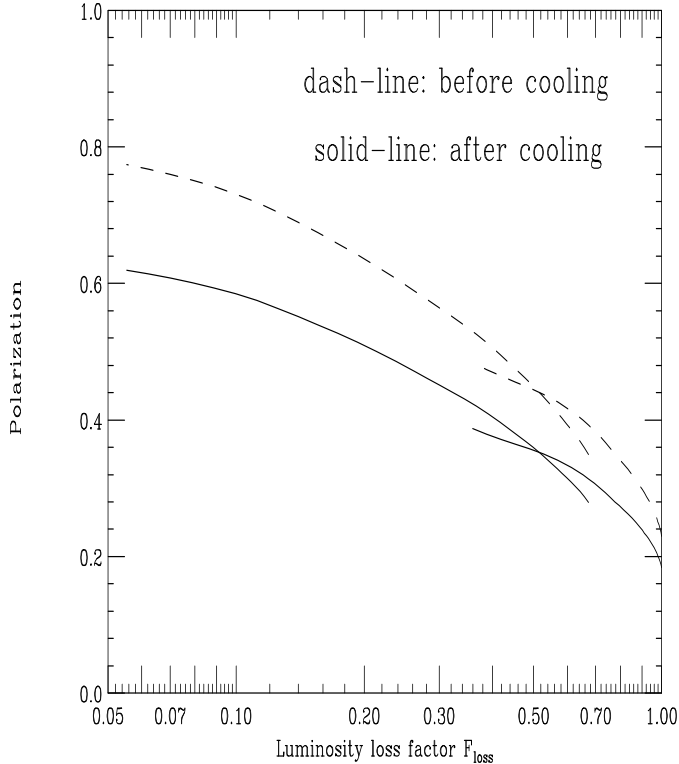,height=4.0in,width=3.5in}}
\caption{Polarization vs $F_{\rm loss}$ of muons 
accepted; 
the dashed line shows 
polarization as selected before cooling; the solid line 
gives polarization 
after cooling.} 
 \label{polvscut}
 \end{figure}

If nothing is done then the average 
muon polarization is about 
0.19. If higher polarization is desired, some selection of 
muons from forward pion decays  $(\cos{\theta_d} 
\rightarrow 1)$ is required. This can be done by momentum 
selecting the muons at the end of the decay and phase 
rotation channel. A snake\cite{drift} is used to generate 
the required dispersion. Varying the selected minimum 
momentum of the muons yields polarization as a function of 
luminosity loss as shown in Fig.\ref{polvscut}. Dilutions 
introduced in the cooling have been calculated\cite{rose} 
and are included. A siberian snake is also required in the 
final collider ring. 

\subsection*{Ionization Cooling}

For the required collider luminosity, the phase-space 
volume must be greatly reduced; and this must be done 
within the $\mu$ lifetime. Cooling by synchrotron 
radiation, conventional stochastic cooling and conventional 
electron cooling are all too slow. Optical stochastic 
cooling\cite{ref11}, electron cooling in a plasma 
discharge\cite{ref12} and cooling in a crystal 
lattice\cite{ref13} are being studied, but appear very 
difficult. Ionization cooling\cite{ref14} of muons seems 
relatively straightforward. 

In ionization cooling, the beam loses both transverse and 
longitudinal momentum as it passes through a material 
medium. Subsequently, the longitudinal momentum can be 
restored by coherent reacceleration, leaving a net loss of 
transverse momentum. 

The equation for transverse cooling (with energies in 
GeV)  is:
  \begin{equation}
\frac{d\epsilon_n}{ds}\ =\ -\frac{dE_{\mu}}{ds}\ 
\frac{\epsilon_n}{E_{\mu}}\ +
\ \frac{\beta_{\perp} (0.014)^2}{2\ E_{\mu}m_{\mu}\ 
L_R},\label{eq1}
  \end{equation}
where $\epsilon_n$ is the normalized emittance, 
$\beta_{\perp}$ is the betatron function at the absorber, 
$dE_{\mu}/ds$ is the energy loss, and $L_R$  is the 
radiation length of the material.  The first term in this 
equation is the coherent cooling term, and the second is 
the heating due to multiple scattering. This heating term 
is minimized if $\beta_{\perp}$ is small (strong-focusing) 
and $L_R$ is large (a low-Z absorber).

Energy spread is reduced by placing a transverse variation 
in absorber density or thickness at a location where 
position is energy dependent, i.e. where there is 
dispersion. The use of such wedges can reduce energy 
spread, but it simultaneously increases transverse 
emittance in the direction of the dispersion. It thus 
allows the exchange of emittance between the longitudinal 
and transverse directions.

\subsubsection{Cooling System}

The cooling is obtained in a series of cooling stages. In 
general, each stage consists of three components with 
matching sections between them: 

 \begin{enumerate}
 \item a FOFO lattice consisting of spaced axial 
solenoids with alternating 
field directions and lithium hydride absorbers placed at 
the centers of the 
spaces between them, where the $\beta_{\perp}$'s are 
minimum. 
 \item a lattice consisting of more widely separated 
alternating solenoids,
and bending magnets between them
to generate dispersion. At the location of maximum
dispersion, wedges of lithium hydride are introduced to 
interchange
longitudinal and transverse emittance.
 \item a linac to restore the energy lost in the 
absorbers.
 \end{enumerate}

  In a few of the later stages, current carrying lithium 
rods replace item (1) above. In this case the rod serves 
simultaneously to maintain the low $\beta_{\perp}$, and 
attenuate the beam momenta. Similar lithium rods, with 
surface fields of $10\,$T , were developed at Novosibirsk 
and have been used as focusing elements at FNAL and 
CERN\cite{ref16}.

\begin{figure}[b!]                     
\centerline{\epsfig{file=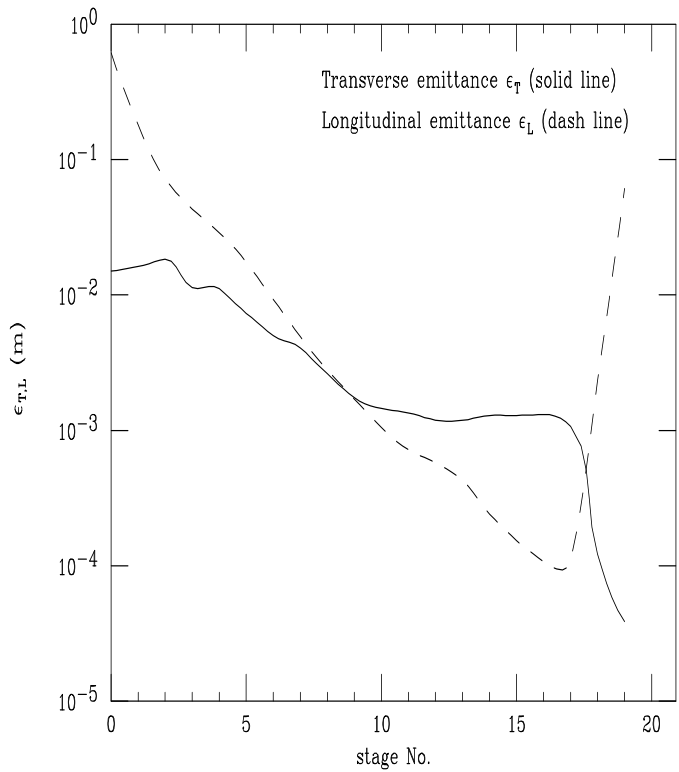,height=4.0in,width=3.5in}}
\caption{Normalized transverse and longitudinal 
emittances as a function of section number in the model 
cooling system}
\label{cooling}
\end{figure}

The emittances, transverse and longitudinal, as a function 
of stage number, are shown in Fig.\ref{cooling}. In the 
first 10 stages, relatively strong wedges are used to 
rapidly reduce the longitudinal emittance, while the 
transverse emittance is reduced relatively slowly. The 
object is to reduce the bunch length, thus allowing the use 
of higher frequency and higher gradient rf in the 
reacceleration linacs. In the next 7 stages, the 
emittances are reduced close to their asymptotic limits. In 
the last 3 stages, using lithium rods, there are no wedges 
and the energy is allowed to fall to about 15 MeV. 
Transverse cooling continues, and the momentum spread is 
allowed to rise. 
   The total length of the system is 750 m, and the total 
acceleration used is 5 GeV. The fraction of muons remaining 
at the end of the cooling system is calculated to be 
$55\,$\%. 

\subsection{Acceleration}
Following cooling and initial bunch compression the 
beams must be rapidly 
accelerated
to full energy (2 TeV, or 250 GeV).  A sequence of 
recirculating accelerators 
(similar to that used at CEBAF)could be used but would 
be relatively 
expensive. A more economical solution would be to use 
fast pulsed magnets in 
synchrotrons with rf systems consisting of significant 
lengths of 
superconducting linac.

For the final acceleration to 2 TeV in the high energy 
machine, the power 
consumed by a ring using only pulsed magnets would be 
excessive and 
alternating pulsed and superconducting 
magnets\cite{summers} are used 
instead.

\subsection*{Collider Storage Ring} 

After acceleration, the $\mu^+$ and $\mu^-$ bunches are 
injected into a 
separate storage ring. The highest possible average 
bending field is 
desirable to maximize the number of revolutions before 
decay, and thus 
maximize the luminosity. Collisions occur in one, or 
perhaps two, very 
low-$\beta^*$ interaction areas. 

\subsubsection*{Bending Magnet Design}

The magnet design is complicated by the fact that the 
$\mu$'s decay within
the rings ($\mu^-\ \rightarrow\ 
e^-\overline{\nu_e}\nu_{\mu}$), producing
electrons whose mean energy is approximately 0.35 that 
of the muons. These
electrons travel toward the inside of the ring dipoles, 
radiating a 
fraction of their energy as synchrotron radiation 
towards the outside of the
ring, and depositing the rest on the inside.  The total 
average power deposited,
in the ring, in the 4 TeV machine is 13 MW.
The beam must thus be surrounded by a $\approx$ 6 cm 
thick warm
shield\cite{iuliupipe}, 
which is
located inside a large aperture conventional 
superconducting magnet.

The quadrupoles can use warm iron 
poles placed as close to the beam as practical, with 
coils either 
superconducting or warm, as dictated by cost 
considerations. 

\subsubsection{Lattice}

In order to maintain a bunch with rms length 3 mm, 
without excessive rf, an 
isochronous lattice, of the dispersion wave 
type\cite{ref17} is used. For the 
3 mm beta at the intersection point, the maximum beta's 
in both x and y are of 
the order of 400 km (14 km in the 0.5 TeV machine). 
Local chromatic 
correction is essential. 
Two lattices have been 
generated\cite{garrenlat}\cite{oidelat}, one of 
which\cite{oidelat}, after the application of octupole 
and decapole correctors, 
has been shown to have an adequate calculated dynamic 
aperture. 

Studies of the  resistive wall impedance instabilities 
indicate that the required muon bunches would be 
unstable if uncorrected. 
In any case, the rf requirements to maintain such 
bunches 
would be excessive. BNS\cite{bns} damping, applied by rf 
quadrupoles\cite{chaobook}, is one 
possible solution, but needs more careful study.

\subsection*{Muon Decay Background}
Monte Carlo study\cite{ref20},\cite{iuliupipe} indicated 
that  the 
background, though serious, should not be impossible. 
Further reductions 
are expected as the shielding is optimized, and it 
should 
be possible to design detectors that are less sensitive 
to the neutrons and 
photons present. 

   There would also be a background from the presence of 
a halo of near full energy muons in the circulating beam. 
The beam will need careful preparation before injection 
into the collider, and a collimation system will have to be 
designed to be located on the opposite side of the ring 
from the detector. 

   There is a small background from incoherent (i.e. \mumu 
$\rightarrow$ \ee) pair production in the 4 TeV Collider 
case. The cross section is estimated to be $10\ mb$, which 
would give rise to a background of $\approx 3\,10^4$  
electron pairs per bunch crossing. Approximately $90\,\%$ 
of these, will be trapped inside the tungsten nose cone, 
but those with energy between 30 and $100\,$MeV will enter 
the detector region.

\section{Conclusion}

 \begin{itemize}
 \item Considerable progress has been made on a scenario 
for a 2 + 2 TeV, high 
luminosity collider. Much work remains to be done, but 
no obvious show stopper 
has yet been found. 
 \item The two areas that could present serious problems 
are: 1) unforeseen 
losses during the 25 stages of cooling (a 3\% loss per 
stage would be very 
serious); and 2) the excessive detector background from 
muon beam halo. 
 \item Many technical components require development: a 
large high field solenoid for capture, low frequency rf 
linacs, multi-beam pulsed and/or rotating magnets for 
acceleration, warm bore shielding inside high field dipoles 
for the collider, muon collimators and background shields, 
etc.\ but: 
 \item None of the required components may be described as 
{\it exotic}, and their specifications are not far beyond 
what has been demonstrated. 
 \item If the components can be developed and the problems 
overcome, then a muon-muon collider could be a 
useful complement to \ee colliders, and, at higher energies 
could be  a viable alternative. 
 \end{itemize}

\section{Acknowledgment}
%
This research was supported by the U.S. Department of 
Energy under Contract No.
DE-ACO2-76-CH00016 and DE-AC03-76SF00515.

 \end{document}